\documentclass[prd,twocolumn,preprintnumbers,nofootinbib]{revtex4}
\pdfoutput=1
\usepackage[a4paper, hdivide={1.91cm,,1.165cm}, vdivide={1.83cm,,2.2cm}]{geometry}

\usepackage{amsmath,amssymb}
\usepackage{graphicx,multirow}
\usepackage{slashed}
\usepackage{units}
\usepackage[hyperfootnotes=false]{hyperref}

\def \L {\mathcal{L}} 
\newcommand{\hc}{\ensuremath{\text{h.c.}}}
\newcommand{\BR}{\ensuremath{\text{BR}}}

\def \i {\mathrm{i}\mkern1mu} 

\allowdisplaybreaks

\begin{document}

\title{Light particles with baryon and lepton numbers}

\author{Julian Heeck }
\email{heeck@virginia.edu}
\affiliation{Department of Physics, University of Virginia, Charlottesville, Virginia 22904, USA}

\begin{abstract}
We consider light new particles $\chi$ and $\phi$ that carry baryon and lepton numbers. If these particles are lighter than nucleons they lead to exotic decays such as $p\to \pi^+ \chi$ and $p\to e^+\phi$, not yet fully constrained by dedicated searches. For $\chi$ and $\phi$ masses in the GeV range, proton decays are kinematically forbidden but other decays of the forms baryon\,$\to$\,meson$+\chi$, meson\,$\to$\,baryon$+\bar\chi$, and baryon\,$\to$\,anti-lepton$+\phi$ involving heavy initial hadrons are allowed. This opens up the possibility to search for apparent baryon number violation not just in underground experiments such as Super-Kamiokande and DUNE but also in decays of heavy hadrons in charm and $B$ factories.
\end{abstract}

\maketitle


\section{Introduction}

There is growing interest in the particle-physics community in light neutral particles that may have evaded our experiments due to their small couplings. The simplest examples of light new scalars, fermions, and vector bosons have been studied extensively in the literature, both in UV-complete realizations and in the framework of effective field theories, see Ref.~\cite{Beacham:2019nyx} for a recent review. In most studies it is assumed that these new light particles do not carry any conserved quantum numbers, which is needlessly restrictive and misses interesting models and signatures.

Seeing as all known fermions carry globally-conserved baryon or lepton numbers, we will discuss scenarios in which new light particles also carry $B$ or $L$~\cite{Heeck:2019ego}, which leads to experimental signatures seldom discussed in the literature. The simplest $U(1)_B$ and $U(1)_L$ assignments are as follows:
\begin{enumerate}
\item Dirac fermion $N$ with $L=1$. This allows for a renormalizable coupling $\overline{L}H N$ and leads to $N$ mixing with neutrinos. Such sterile (Dirac) neutrinos have been discussed extensively in the literature, see e.g.~Ref.~\cite{Bolton:2019pcu} for a recent collection of constraints.	
\item Dirac fermion $\chi$ with $B=1$ and effective coupling to e.g.~$udd\bar{\chi}/\Lambda^2$. This coupling induces a mixing of $\chi$ with neutral baryons such as the neutron and has been discussed in 
Refs.~\cite{Thomas:1995ze,Kitano:2008tk,Davoudiasl:2010am,Allahverdi:2013mza,Davoudiasl:2013pda,Davoudiasl:2014gfa,McKeen:2015cuz,Fornal:2018eol,McKeen:2018xwc,Helo:2018bgb,Karananas:2018goc,Jin:2018moh,Berezhiani:2018udo,McKeen:2020zni}, motivated in part by the neutron lifetime anomaly~\cite{Fornal:2020gto}.
\item Complex scalar $\delta$ with $L=2$ and effective coupling to the Weinberg operator $(\overline{L}H)^2\delta/\Lambda^2$. This leads to an interaction vertex $\bar{\nu}\bar{\nu}\delta$ and has been discussed in Refs.~\cite{Berryman:2018ogk,deGouvea:2019qaz}.
\item Complex scalar $\phi$ with $L=B=1$ and effective coupling to e.g.~$QQQL\bar{\phi}/\Lambda^3$, leading to low-energy interaction vertices such as $p\ell\bar{\phi}$~\cite{McKeen:2020zni}.
\item Complex scalar $\xi$ with $B=2$. An effective coupling to e.g.~$(udd)^2\bar{\xi}/\Lambda^6$ leads to an interaction vertex $n n \bar{\xi}$ at low energies.
\end{enumerate}

We restrict ourselves to these simplest cases because particles with larger $B$ or $L$ have effective couplings that are very suppressed  and will be difficult to observe outside of nuclear decays or indirect dark-matter detection~\cite{Heeck:2019ego}.
If the scalars above obtained vacuum expectation values, the effective operators would lead to neutrinoless double beta decay ($\delta$), proton decay ($\phi$), and neutron--anti-neutron oscillation ($\xi$), all of which are well covered in the literature. 
In this article we are interested in the case of unbroken global (or even local~\cite{Heeck:2014zfa}) $U(1)_{B,L}$ symmetries, so we assume scalar potentials that do not lead to vacuum expectation values for these new scalars. This opens up novel signatures of \emph{apparent} baryon and lepton number violation in heavy hadron decays.

\section{Fermion with \texorpdfstring{$B=1$}{B=1}}

New fermions carrying lepton number $L=1$, i.e.~sterile neutrinos, have been widely discussed in the literature. We will therefore focus on the \emph{sterile neutron} case: a new fermion $\chi$ with baryon number $B(\chi)=1$ and effective Lagrangian~\cite{delAguila:2008ir}
\begin{align}
\L_\chi = \bar{\chi} (\i \slashed{\partial}-m_\chi  )\chi+ \left(\frac{u_i d_j d_k \chi_L^c}{\Lambda_{ijk}^2}+ \frac{Q_i Q_j d_k \chi_L^c}{\tilde\Lambda_{ijk}^2} +\hc\right) .
\label{eq:lagrangian_chi}
\end{align}
Here and in the following we only show the flavor indices and suppress Lorentz, color, and isospin indices. Since we assume baryon number to be conserved, this Lagrangian does not lead to neutron--anti-neutron oscillations. The new neutral fermion $\chi$ will, however, eventually mix with the neutron (and all other neutral baryons) and inherit all of the neutron's interactions, allowing us to produce and detect it.
Simple UV completions for this model can be found in Refs.~\cite{Fornal:2018eol,Jin:2018moh}. For simplicity we will focus on the interaction term $udd$ involving right-handed quarks in the following, since the $QQd$ term yields essentially the same phenomenology.
 
The phenomenology of $\chi$ depends crucially on its mass: if $\chi$ is the lightest baryon,  it will induce proton and neutron decays that are stringently constrained by experiments such as Super-Kamiokande~\cite{Heeck:2019kgr}; however, if $\chi$ is \emph{heavier} than the neutron, nuclei will remain stable and $\chi$ could still be produced in decays of heavier baryons and mesons.

\subsection{\texorpdfstring{$\chi$}{chi} as the lightest neutral baryon}

If $\chi$ is lighter than the neutron, the above Lagrangian will generate the so-far unobserved neutron decay $n\to \chi \gamma$~\cite{Davoudiasl:2014gfa,Fornal:2018eol}.
For $m_\chi < m_p + m_e$, the hydrogen decay  $H\to \chi \nu$~\cite{Berezhiani:2018udo,McKeen:2020zni} also opens up, and for $m_\chi < m_p-m_e$ the proton decay channel  $p\to \chi e^+\nu$~\cite{McKeen:2020zni}. 
Lowering the $\chi$ mass further eventually allows for search-friendly two-body nucleon decay channels such as $n\to \pi^0\chi$ and $p\to \pi^+\chi$~\cite{Davoudiasl:2013pda,Davoudiasl:2014gfa,Helo:2018bgb}, with a decay rate of the order
\begin{align}
\begin{split}
\Gamma (p\to \chi \pi^+) &\simeq \frac{|\vec{p}_\pi|}{16\pi}\, \frac{W_0^2}{\Lambda_{udd}^4}\left(1+\frac{m_\chi^2}{m_p^2}-\frac{m_\pi^2}{m_p^2}\right)\\
& \sim \frac{1}{\unit[10^{33}]{yr}}\left(\frac{\unit[2\times 10^{15}]{GeV}}{\Lambda_{udd}}\right)^4 ,
\end{split}
\label{eq:proton_to_pion}
\end{align}
where $W_0\simeq \unit[0.189]{GeV^2}$~\cite{Aoki:2017puj} describes the matrix element $\langle \pi^+ | (ud)_R d_R|p\rangle$.
Limits on these nucleon decay modes only exist for $m_\chi\ll m_p$, where $\chi$ mimics a neutrino: $\tau (p\to \pi^+ \nu) > \unit[390\times 10^{30}]{yr}$~\cite{Abe:2013lua}, $\tau (n\to \pi^0 \nu) > \unit[1100\times 10^{30}]{yr}$~\cite{Abe:2013lua}, $\tau (p\to e^+ \nu\nu) > \unit[170\times 10^{30}]{yr}$~\cite{Takhistov:2014pfw}.
These limits push the scale $\Lambda_{udd}$ above $\unit[10^{15}]{GeV}$, similar to other proton decay operators. Limits of similar order can be expected for larger $m_\chi$ as long as the emitted charged pion is above the Cherenkov threshold~\cite{Heeck:2019kgr}.

The coupling $\Lambda_{uds}$ in Eq.~\eqref{eq:lagrangian_chi} will instead lead to $n\to K^0\chi$ and $p\to K^+\chi$~\cite{Davoudiasl:2013pda,Davoudiasl:2014gfa}, the latter being even better constrained than the pion mode for $m_\chi \ll m_p$~\cite{Abe:2014mwa}. Since the outgoing kaon in $p\to K^+\nu$ is already below the Cherenkov threshold its momentum is not important in the analysis, so the limit of Ref.~\cite{Abe:2014mwa} should apply to all $p\to K^+\chi$ independent of $m_\chi$.

Naively it seems possible to eliminate nucleon decays even for $m_\chi < m_n$ by considering coupling constants $\Lambda_{ijk}$ with second or third-generation quarks, too heavy to be produced. However, at loop level even a third-generation coupling such as $\Lambda_{tbb}$ will induce $p\to \pi^+ \chi$, $p\to K^+\chi$, and $n\to \gamma \chi$, which will hence remain the best signatures of all $\Lambda_{ijk}$ for $m_\chi < m_n$. We strongly encourage dedicated searches for these two-body nucleon decay modes for the entire mass range $0\leq m_\chi< m_n$ in Super-Kamiokande as well as in the future facilities JUNO, DUNE, and Hyper-Kamiokande.

Notice that $\chi$ is stable in the mass regime $m_\chi< m_p+m_e$~\cite{Allahverdi:2013mza} and will therefore form dark matter~\cite{Fornal:2020bzz}. If the interactions underlying $\L_\chi$ are in equilibrium in the early universe we expect $\chi$ be \emph{asymmetric} dark matter~\cite{Zurek:2013wia} on account of its baryon number. Additional interactions are typically necessary to produce the correct abundance, which will however not influence the processes of interest here. 

\subsection{\texorpdfstring{$\chi$}{chi} heavier than the neutron}

Proton and neutron decays are kinematically forbidden for $m_\chi > m_n$, which drastically changes the phenomenology of $\chi$. For $m_\chi$ in the GeV range, $\chi$ can still be produced hadron decays, notably the two-body decays
\begin{align}
\begin{split}
\text{baryon}\to \text{meson} + \chi\,,\\
\text{meson}\to \text{baryon} + \bar{\chi}\,,
\end{split}
\label{eq:chi_processes}
\end{align}
where $\chi$ typically leaves the detector as missing energy.\footnote{For $\Lambda < \unit{TeV}$ a displaced-vertex signature from $\chi$ decays such as $\chi\to p\pi^-$ is possible as well, but will not be discussed here.}
To produce the heavy $\chi$ the initial hadron and the interaction Lagrangian must contain second or third-generation quarks.
Such decays were already proposed and discussed in Refs.~\cite{Aitken:2017wie,Elor:2018twp}.
Other decays of interest such as $\text{baryon}^0\to \gamma + \chi$ and $\text{baryon}^+\to \ell^++\nu + \chi$ are suppressed by $\alpha_\text{EM}$ and $G_F$, respectively, compared to pure hadronic decays.

The calculation of the decay rates in Eq.~\eqref{eq:chi_processes} via the Lagrangian $\L_\chi$ is difficult because the relevant baryon-number-violating matrix elements $\langle \text{baryon} | qqq |\text{meson}\rangle$ have, to the best of our knowledge, not been calculated for the charm and bottom quarks of interest here. We will resort to simple order-of-magnitude estimates for now and leave a more sophisticated calculation for future work. For the decays $\text{baryon}\to \text{meson} + \chi $ we will use Eq.~\eqref{eq:proton_to_pion} with the appropriate replacement of masses and indices, keeping $W_0\sim \unit[0.1]{GeV^2}$ as an appropriate QCD matrix element independent of the quark flavors.
For meson decay to baryons we perform an analogous calculation to obtain
\begin{align}
\Gamma ( \text{mes}\to \text{bar} + \bar{\chi}) \simeq \frac{W_0^2 |\vec{p}_{\chi}|}{16\pi \Lambda^4}\left( 1 -\frac{m_\chi^2+m_\text{bar}^2}{m_\text{mes}^2}\right), 
\end{align}
with appropriate indices for $\Lambda$ and again $W_0\sim \unit[0.1]{GeV^2}$.

\begin{figure*}[t]
\includegraphics[width=0.70\textwidth]{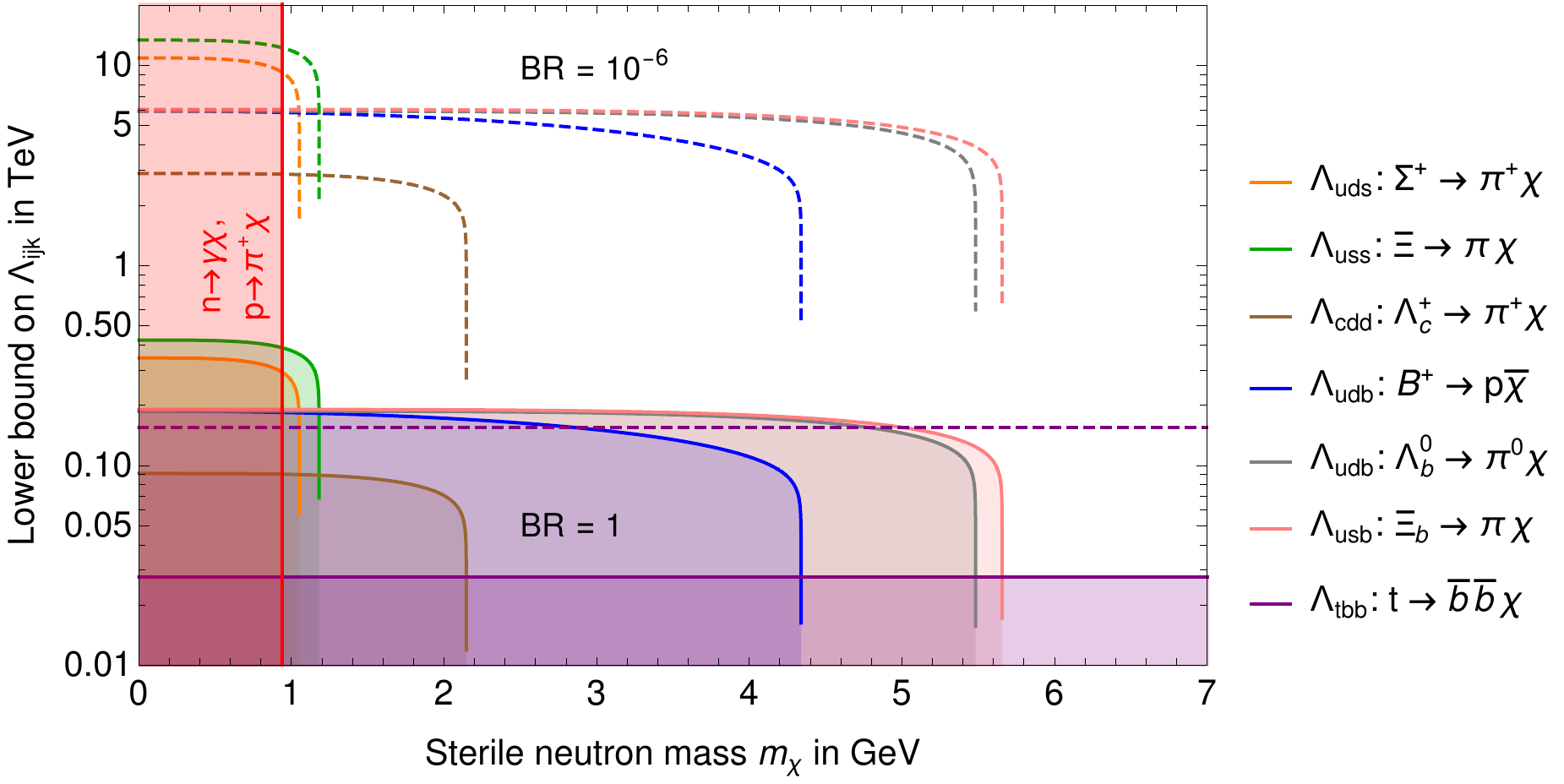}
\caption{Lower bounds on the suppression scale $\Lambda_{ijk}$ in $\L_\chi$ from various rare hadron decays. The shaded regions are excluded using lifetime measurements, demanding the new decay channels to have branching ratios below $1$. The dashed lines show constraints if the decays involving $\chi$ could be probed to branching ratios below $10^{-6}$ in dedicated searches. See text for details.
}
\label{fig:sterile_neutron_bounds}
\end{figure*}

A by no means exhaustive collection of constraints is given in Fig.~\ref{fig:sterile_neutron_bounds}. Since none of the processes in Eq.~\eqref{eq:chi_processes} have been experimentally searched for so far, we only show the most basic constraints from demanding that these decays have a branching ratio below one. This translates into lower bounds on $\Lambda$ below $\unit[400]{GeV}$; new-physics scales this low are likely better constrained by LHC searches for the underlying colored mediator particles~\cite{Elor:2018twp}. However, dedicated searches can improve the rare-decay limits considerably:
\begin{itemize}
\item Samples of up to a million $\Sigma^+$ have been used 40 to 50 years ago to measure  $\Sigma^+\to n\pi^+$~\cite{Marraffino:1980dj,Zyla:2020zbs} and constrain $\Sigma^+\to n \ell^+\nu$~\cite{Eisele:1969fz}. Revisiting this data should make it possible to constrain $\Sigma^+\to \pi^+\chi$ to branching ratios around $10^{-5}$ for $m_\chi\sim m_n$. $\Sigma^+$ are also copiously produced at hadron colliders and LHCb has already collected more than $10^{14}$~\cite{Aaij:2017ddf}; it should be possible to obtain excellent limits beyond $10^{-6}$ using this enormous dataset even though the decay $\Sigma^+\to \pi^+\chi$ is not particularly clean in LHCb if $\chi$ leaves the detector.
\item $\Xi^-\to n\pi^-$ has been constrained to branching ratios $2\times 10^{-5}$ about 40 years ago~\cite{Biagi:1982eu} and similar limits should apply to $\Xi^-\to \chi\pi^-$ with $m_\chi\sim m_n$. A much larger dataset of around $10^9$ $\Xi^-$ has been collected in HyperCP~\cite{Rajaram:2005bs} and should make it possible to improve this limit further down to  $10^{-6}$.
\item $\Lambda_c^+$ are abundantly produced at Belle II, BESIII, and LHCb, which should be able to push $\BR(\Lambda_c^+\to \pi^+\chi)\ll 1$. The proposed Super Tau Charm Facility could also perform this and other searches~\cite{Shi:2019vus}.
\item The $B$-factories Belle and BaBar have collected data of roughly $10^9$ $B$ meson decays, with Belle II aiming to collect $5\times 10^{10}$. Decays such as $B^+ \to p\bar{\chi}$  are sufficiently clean to be probed down to tiny branching ratios, at least $10^{-6}$, most likely lower.
\item Decays of $\Lambda_b^0$ and $\Xi_b$ into $\pi^0\chi$ can reach $\chi$ masses above $\unit[5]{GeV}$ and could once again be probed in $B$ factories.
\item For $m_\chi$ exceeding $\sim\unit[6]{GeV}$, only top-quark decays have enough energy to produce $\chi$ on shell. Judging by existing constraints on the $\Delta B =1$ top-quark decays $\BR (t\to \ell+2\text{jets}) < 0.0017$~\cite{Chatrchyan:2013bba}, it seems unlikely that our decays such as $t\to\bar b\bar b \chi$ could be probed far below $10^{-3}$. As a result, top quark decays will probe effective scales $\Lambda$ below $\unit[100]{GeV}$, where our effective field theory breaks down and should be replaced by a UV-complete model.
\end{itemize}
Above and in Fig.~\ref{fig:sterile_neutron_bounds} we have focused only on a small subset of possible decay channels.
Although dedicated analyses have to be performed in all cases in order to assess the actual experimental sensitivity to a given channel, it seems feasible to reach branching ratio limits below $10^{-6}$ for $\unit[1]{GeV}\lesssim m_\chi\lesssim \unit[5]{GeV}$ with existing experiments.
This then probes effective couplings $\Lambda$ in the multi-TeV region, completely uncharted terrain!

The reader might be worried about LHC constraints on the underlying scalar di-quarks that generate the couplings in Eq.~\eqref{eq:lagrangian_chi}. Resonantly produced via $qq\to\,$di-quark these are constrained to be heavier than $\unit[7]{TeV}$ by CMS~\cite{Sirunyan:2018xlo}.
In our case, however, the couplings of interest for Fig.~\ref{fig:sterile_neutron_bounds} couple the di-quark to at least one second or third generation quark, e.g.~to $u b$. This coupling precludes an efficient resonant production at the LHC and is thus not constrained by existing searches. Furthermore, the di-quark decay channel could be dominantly into $q \chi$ rather than $qq$, which would suppress the search efficiency even more. Dedicated direct collider searches for the underlying di-quarks are necessary to provide complementary constraints to the heavy hadron decays of Fig.~\ref{fig:sterile_neutron_bounds}.

\section{Scalar with \texorpdfstring{$B=1$}{B=1} and \texorpdfstring{$L=1$}{L=1}}

\begin{figure}[b]
\includegraphics[width=0.4\textwidth]{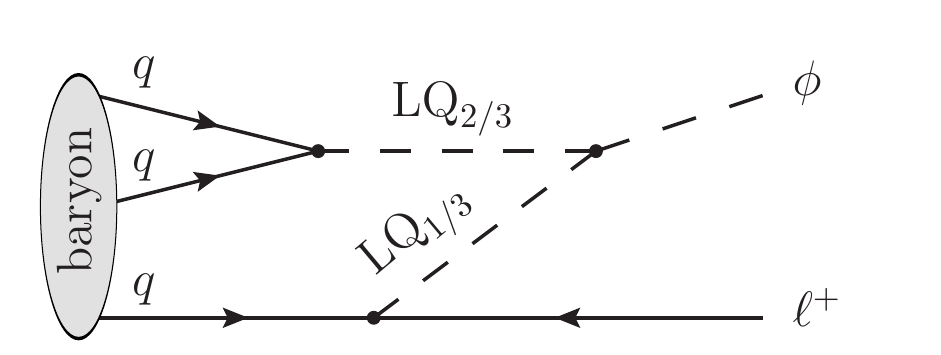}
\caption{UV-complete realization of the decay baryon$\to \ell^+\phi$. LQ$_B$ are two  heavy leptoquarks~\cite{Dorsner:2016wpm} with the same gauge quantum numbers but different baryon numbers~$B$. Since we assume baryon number to be conserved in this model all baryon decays contain $\phi$ in the final state.
}
\label{fig:sterile_leptoquark}
\end{figure}

New singlet scalars carrying lepton number $L=2$ have been discussed recently in Refs.~\cite{Berryman:2018ogk,deGouvea:2019qaz}. Here we focus on the case of scalars $\phi$ with $B(\phi)=L(\phi)=1$ and effective Lagrangian
\begin{align}
\L_\phi = & |\partial\phi|^2 - m_\phi^2 |\phi|^2+ \left(
\frac{d_i u_j u_k \ell_l \phi^*}{\Lambda_{ijkl}^3} +
\frac{d_i u_j Q_k L_l \phi^*}{\tilde\Lambda_{ijkl}^3} \right.\nonumber\\
&\left. +
\frac{Q_i Q_j u_k \ell_l \phi^*}{\Lambda_{ijkl}'^3} +
\frac{Q_i Q_j Q_k L_l \phi^*}{\tilde\Lambda_{ijkl}'^3} +\hc\right) ,
\label{eq:lagrangian_phi}
\end{align}
ignoring couplings in the scalar potential. Since $\phi$ has no gauge interactions but still carries baryon and lepton number it could be called a \emph{sterile leptoquark}. UV completions of these operators are easy to construct and sketched in Fig.~\ref{fig:sterile_leptoquark}.
Additional lepton \emph{flavor} symmetries could be imposed in order to restrict the couplings~\cite{Heeck:2016xwg,Hambye:2017qix} but will not be discussed here. 
Just like in the case of the sterile neutron above the phenomenology differs drastically depending on $\phi$'s mass relative to $m_n$.

\subsection{\texorpdfstring{$\phi$}{phi} lighter than the neutron}

For $m_\phi<m_n$, the Lagrangian $\L_\phi$ will lead to the neutron decay $n\to \phi\bar{\nu}$, no matter the underlying operator or flavor structure. Limits of order $\unit[10^{31}]{yr}$ on such an invisible neutron decay have been obtained in KamLAND~\cite{Araki:2005jt} and can be improved with future detectors~\cite{Heeck:2019kgr}.
Other models leading to invisible neutron decay have been discussed in Ref.~\cite{Barducci:2018rlx}.
For $m_\phi < m_p + m_e$, the hydrogen channel $H\to \phi\gamma$ opens up and for $m_\phi <m_p - m_e$ the two-body proton decay channel $p\to e^+\phi$~\cite{McKeen:2020zni}:
\begin{align}
\begin{split}
\Gamma (p\to e^+\phi) &\simeq \frac{|\vec{p}_e|}{16\pi}\, \frac{\beta^2}{\Lambda_{duue}^6}\left(1+\frac{m_e^2}{m_p^2}-\frac{m_\phi^2}{m_p^2}\right)\\
& \sim \frac{1}{\unit[10^{33}]{yr}}\left(\frac{\unit[7\times 10^{9}]{GeV}}{\Lambda_{duue}}\right)^6 ,
\end{split}
\label{eq:proton_to_electron}
\end{align}
where $\beta\simeq \unit[0.014]{GeV^3}$~\cite{Aoki:2017puj} describes the QCD matrix element $\langle 0 | (ud)_R u_R|p\rangle$.

Limits on $p\to e^+\phi$ and $p\to \mu^+\phi$ for $m_\phi=0$ have been obtained in Super-Kamiokande and exclude lifetimes below $\unit[8\times 10^{32}]{yr}$ and $\unit[4\times 10^{32}]{yr}$~\cite{Takhistov:2015fao}, respectively.
$\L_\phi$ terms involving second and third-generation quarks will also induce the two-body decays from above at loop level, rendering them the cleanest probe for $m_\phi < m_n$. Notice that $\phi$ is stable for $m_\phi < m_p+m_e$ and will thus form dark matter, albeit with an abundance that depends on a variety of other factors.

\subsection{\texorpdfstring{$\phi$}{phi} heavier than the neutron}

\begin{figure}[t]
\includegraphics[width=0.4\textwidth]{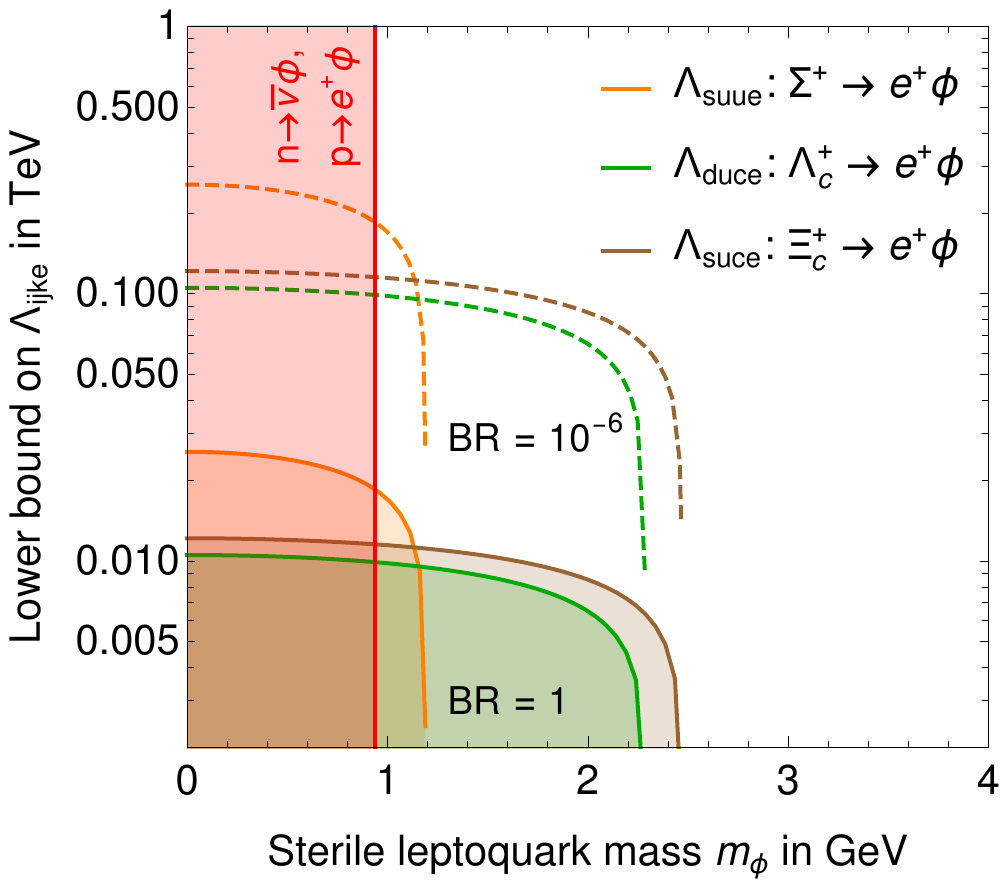}
\caption{Lower bounds on the suppression scale $\Lambda_{ijke}$ from various rare baryon decays. The shaded regions are excluded using lifetime measurements, the dashed lines show constraints if the decays involving $\phi$ could be probed to branching ratios below $10^{-6}$ in dedicated searches. See text for details.
}
\label{fig:sterile_leptoquark_bounds}
\end{figure}

The stringently constrained nucleon decays involving $\phi$ are absent for $m_\phi > m_n$, but $\L_\phi$ still allows for two-body decays of the form
\begin{align}
\begin{split}
\text{baryon}\to \ell^+ + \phi\,.
\end{split}
\label{eq:phi_processes}
\end{align}
Notice that decays $\tau^+\to \text{baryon} + \bar{\phi}$ are kinematically forbidden for $m_\phi > m_n$.
Using Eq.~\eqref{eq:proton_to_electron} with the appropriate replacement of labels and indices -- but keeping the same QCD matrix element for lack of dedicated calculations -- we can calculate the decay rates of Eq.~\eqref{eq:phi_processes} for some baryons of interest, shown in Fig.~\ref{fig:sterile_leptoquark_bounds}.
Since the relevant operators in $\L_\phi$ are of mass dimension 7, the decay rates are suppressed by an additional factor $(\Lambda_\text{QCD}/\Lambda)^2$ compared to the processes involving sterile neutrons. This results in rather weak limits on $\Lambda_{duu\ell}$ that are typically better constrained by LHC searches for the underlying heavy mediators. Pushing $\Lambda_{duu\ell}$ above TeV requires measurements of branching ratios around $10^{-12}$, which are difficult to achieve for the processes of interest here despite the large number of heavy baryons in e.g.~LHCb~\cite{Aaij:2017ddf}

\section{Scalar with \texorpdfstring{$B=2$}{B=2}}

Let us briefly discuss the last case of potential interest, a new scalar $\xi$ with baryon number $B=2$ and effective couplings $ udd udd \bar{\xi}/\Lambda^6$. The high mass dimension of this operator makes it difficult to induce observable rates in heavy hadron decays such as baryon\,$\to$\,$\xi$\,$+$\,anti-baryon. The only realistic signatures to probe this model are nuclear decays, e.g.~$n n \to \xi \pi^0$ or the fully invisible $(A,Z)\to (A-2,Z)+\xi$ for $m_\xi \lesssim 2m_n$. The invisible channel leads to de-excitation radiation as the daughter nucleus relaxes to its ground state and has been constrained in KamLAND to lifetimes above $\unit[10^{30}]{yr}$~\cite{Araki:2005jt}, to be improved in future facilities~\cite{Heeck:2019kgr}. While a calculation of the decay width is difficult, simple estimates show that this pushes the limit on $\Lambda_{uddudd}$ above the TeV scale.

\section{Summary and Discussion}
\label{sec:conclusions}

The phenomenology of light new particles changes drastically if they carry baryon or lepton numbers. In the simplest examples discussed in this article we find a variety of nuclear decay channels such as $p\to \pi^+\chi$ and $p\to \ell^+\phi$ that have not yet been explored experimentally for non-zero $m_{\chi,\phi}$. Of particular interest are invisible (multi-)neutron decays such as $n\to \bar\nu \phi$ and $nn\to \xi$ because they do not depend on the masses of the new particles. Dedicated searches or coverage via inclusive searches~\cite{Heeck:2019kgr} in Super-Kamiokande and future detectors are required in order not to miss these sub-GeV particles. Particles with masses in the GeV range would not induce proton decay but could still be produced in heavy hadron decays of the forms baryon\,$\to$\,meson$+\chi$, meson\,$\to$\,baryon$+\bar\chi$, and baryon\,$\to$\,anti-lepton$+\phi$. These apparently baryon-number-violating decays can be searched for in charm and $B$ factories such as BESIII, Belle II and LHCb. On the experimental side this requires dedicated sensitivity studies in order to identify the most promising mother and daughter particles; on the theoretical side this requires a calculation of the relevant matrix elements.
Ultimately a more complete model for these particles is desirable that connects them to dark matter and the baryon asymmetry of our universe, to be discussed in future work.

\section*{Acknowledgements}
I thank Arvind Rajaraman, Hooman Davoudiasl, and Simon Knapen for discussions.

\bibliographystyle{utcaps_mod}
\bibliography{BIB}

\providecommand{\href}[2]{#2}\begingroup\raggedright\begin{thebibliography}{10}

\bibitem{Beacham:2019nyx}
J.~Beacham {\em et~al.}, ``{\em {Physics Beyond Colliders at CERN: Beyond the
  Standard Model Working Group Report}},''
  \href{http://dx.doi.org/10.1088/1361-6471/ab4cd2}{J. Phys. G {\bfseries 47}
  (2020) 010501}, \href{http://arxiv.org/abs/1901.09966}{[{\ttfamily
  1901.09966}]}.

\bibitem{Heeck:2019ego}
J.~Heeck and A.~Rajaraman, ``{\em {How to produce antinuclei from dark
  matter}},'' \href{http://dx.doi.org/10.1088/1361-6471/ab9f03}{J. Phys.
  {\bfseries G47} (2020) 105202},
\href{http://arxiv.org/abs/1906.01667}{[{\ttfamily 1906.01667}]}.

\bibitem{Bolton:2019pcu}
P.~D. Bolton, F.~F. Deppisch, and P.~S. Bhupal~Dev, ``{\em {Neutrinoless double
  beta decay versus other probes of heavy sterile neutrinos}},''
  \href{http://dx.doi.org/10.1007/JHEP03(2020)170}{JHEP {\bfseries 03} (2020)
  170},
\href{http://arxiv.org/abs/1912.03058}{[{\ttfamily 1912.03058}]}.

\bibitem{Thomas:1995ze}
S.~D. Thomas, ``{\em {Baryons and dark matter from the late decay of a
  supersymmetric condensate}},''
  \href{http://dx.doi.org/10.1016/0370-2693(95)00772-D}{Phys. Lett. B
  {\bfseries 356} (1995) 256--263},
  \href{http://arxiv.org/abs/hep-ph/9506274}{[{\ttfamily hep-ph/9506274}]}.

\bibitem{Kitano:2008tk}
R.~Kitano, H.~Murayama, and M.~Ratz, ``{\em {Unified origin of baryons and dark
  matter}},'' \href{http://dx.doi.org/10.1016/j.physletb.2008.09.049}{Phys.
  Lett. {\bfseries B669} (2008) 145--149},
\href{http://arxiv.org/abs/0807.4313}{[{\ttfamily 0807.4313}]}.

\bibitem{Davoudiasl:2010am}
H.~Davoudiasl, D.~E. Morrissey, K.~Sigurdson, and S.~Tulin, ``{\em
  {Hylogenesis: A Unified Origin for Baryonic Visible Matter and Antibaryonic
  Dark Matter}},''
  \href{http://dx.doi.org/10.1103/PhysRevLett.105.211304}{Phys. Rev. Lett.
  {\bfseries 105} (2010) 211304},
\href{http://arxiv.org/abs/1008.2399}{[{\ttfamily 1008.2399}]}.

\bibitem{Allahverdi:2013mza}
R.~Allahverdi and B.~Dutta, ``{\em {Natural GeV Dark Matter and the Baryon-Dark
  Matter Coincidence Puzzle}},''
  \href{http://dx.doi.org/10.1103/PhysRevD.88.023525}{Phys. Rev. D {\bfseries
  88} (2013) 023525}, \href{http://arxiv.org/abs/1304.0711}{[{\ttfamily
  1304.0711}]}.

\bibitem{Davoudiasl:2013pda}
H.~Davoudiasl, ``{\em {Gravitationally Induced Dark Matter Asymmetry and Dark
  Nucleon Decay}},'' \href{http://dx.doi.org/10.1103/PhysRevD.88.095004}{Phys.
  Rev. {\bfseries D88} (2013) 095004},
\href{http://arxiv.org/abs/1308.3473}{[{\ttfamily 1308.3473}]}.

\bibitem{Davoudiasl:2014gfa}
H.~Davoudiasl, ``{\em {Nucleon Decay into a Dark Sector}},''
  \href{http://dx.doi.org/10.1103/PhysRevLett.114.051802}{Phys. Rev. Lett.
  {\bfseries 114} (2015) 051802},
\href{http://arxiv.org/abs/1409.4823}{[{\ttfamily 1409.4823}]}.

\bibitem{McKeen:2015cuz}
D.~McKeen and A.~E. Nelson, ``{\em {CP Violating Baryon Oscillations}},''
  \href{http://dx.doi.org/10.1103/PhysRevD.94.076002}{Phys. Rev. D {\bfseries
  94} (2016) 076002}, \href{http://arxiv.org/abs/1512.05359}{[{\ttfamily
  1512.05359}]}.

\bibitem{Fornal:2018eol}
B.~Fornal and B.~Grinstein, ``{\em {Dark Matter Interpretation of the Neutron
  Decay Anomaly}},''
  \href{http://dx.doi.org/10.1103/PhysRevLett.120.191801}{Phys. Rev. Lett.
  {\bfseries 120} (2018) 191801},
\href{http://arxiv.org/abs/1801.01124}{[{\ttfamily 1801.01124}]}.

\bibitem{McKeen:2018xwc}
D.~McKeen, A.~E. Nelson, S.~Reddy, and D.~Zhou, ``{\em {Neutron stars exclude
  light dark baryons}},''
  \href{http://dx.doi.org/10.1103/PhysRevLett.121.061802}{Phys. Rev. Lett.
  {\bfseries 121} (2018) 061802},
  \href{http://arxiv.org/abs/1802.08244}{[{\ttfamily 1802.08244}]}.

\bibitem{Helo:2018bgb}
J.~C. Helo, M.~Hirsch, and T.~Ota, ``{\em {Proton decay and light sterile
  neutrinos}},'' \href{http://dx.doi.org/10.1007/JHEP06(2018)047}{JHEP
  {\bfseries 06} (2018) 047},
\href{http://arxiv.org/abs/1803.00035}{[{\ttfamily 1803.00035}]}.

\bibitem{Karananas:2018goc}
G.~K. Karananas and A.~Kassiteridis, ``{\em {Small-scale structure from neutron
  dark decay}},'' \href{http://dx.doi.org/10.1088/1475-7516/2018/09/036}{JCAP
  {\bfseries 09} (2018) 036},
  \href{http://arxiv.org/abs/1805.03656}{[{\ttfamily 1805.03656}]}.

\bibitem{Jin:2018moh}
M.~Jin and Y.~Gao, ``{\em {Nucleon-Light Dark Matter Annihilation through
  Baryon Number Violation}},''
  \href{http://dx.doi.org/10.1103/PhysRevD.98.075026}{Phys. Rev. {\bfseries
  D98} (2018) 075026},
\href{http://arxiv.org/abs/1808.10644}{[{\ttfamily 1808.10644}]}.

\bibitem{Berezhiani:2018udo}
Z.~Berezhiani, ``{\em {Neutron lifetime and dark decay of the neutron and
  hydrogen}},'' \href{http://dx.doi.org/10.31526/LHEP.1.2019.118}{LHEP
  {\bfseries 2} no.~1, (2019) 118},
  \href{http://arxiv.org/abs/1812.11089}{[{\ttfamily 1812.11089}]}.

\bibitem{McKeen:2020zni}
D.~McKeen and M.~Pospelov, ``{\em {How long does the hydrogen atom live?}},''
\href{http://arxiv.org/abs/2003.02270}{[{\ttfamily 2003.02270}]}.

\bibitem{Fornal:2020gto}
B.~Fornal and B.~Grinstein, ``{\em {Neutron's Dark Secret}},''
\href{http://arxiv.org/abs/2007.13931}{[{\ttfamily 2007.13931}]}.

\bibitem{Berryman:2018ogk}
J.~M. Berryman, A.~De~Gouv{\^e}a, K.~J. Kelly, and Y.~Zhang, ``{\em
  {Lepton-Number-Charged Scalars and Neutrino Beamstrahlung}},''
  \href{http://dx.doi.org/10.1103/PhysRevD.97.075030}{Phys. Rev. {\bfseries
  D97} (2018) 075030},
\href{http://arxiv.org/abs/1802.00009}{[{\ttfamily 1802.00009}]}.

\bibitem{deGouvea:2019qaz}
A.~de~Gouv{\^e}a, P.~S.~B. Dev, B.~Dutta, T.~Ghosh, T.~Han, and Y.~Zhang,
  ``{\em {Leptonic Scalars at the LHC}},''
  \href{http://dx.doi.org/10.1007/JHEP07(2020)142}{JHEP {\bfseries 07} (2020)
  142},
\href{http://arxiv.org/abs/1910.01132}{[{\ttfamily 1910.01132}]}.

\bibitem{Heeck:2014zfa}
J.~Heeck, ``{\em {Unbroken $B - L$ symmetry}},''
  \href{http://dx.doi.org/10.1016/j.physletb.2014.10.067}{Phys. Lett.
  {\bfseries B739} (2014) 256--262},
\href{http://arxiv.org/abs/1408.6845}{[{\ttfamily 1408.6845}]}.

\bibitem{delAguila:2008ir}
F.~del Aguila, S.~Bar-Shalom, A.~Soni, and J.~Wudka, ``{\em {Heavy Majorana
  Neutrinos in the Effective Lagrangian Description: Application to Hadron
  Colliders}},'' \href{http://dx.doi.org/10.1016/j.physletb.2008.11.031}{Phys.
  Lett. {\bfseries B670} (2009) 399--402},
\href{http://arxiv.org/abs/0806.0876}{[{\ttfamily 0806.0876}]}.

\bibitem{Heeck:2019kgr}
J.~Heeck and V.~Takhistov, ``{\em {Inclusive Nucleon Decay Searches as a
  Frontier of Baryon Number Violation}},''
  \href{http://dx.doi.org/10.1103/PhysRevD.101.015005}{Phys. Rev. {\bfseries
  D101} (2020) 015005},
\href{http://arxiv.org/abs/1910.07647}{[{\ttfamily 1910.07647}]}.

\bibitem{Aoki:2017puj}
Y.~Aoki, T.~Izubuchi, E.~Shintani, and A.~Soni, ``{\em {Improved lattice
  computation of proton decay matrix elements}},''
  \href{http://dx.doi.org/10.1103/PhysRevD.96.014506}{Phys. Rev. D {\bfseries
  96} (2017) 014506}, \href{http://arxiv.org/abs/1705.01338}{[{\ttfamily
  1705.01338}]}.

\bibitem{Abe:2013lua}
{\bfseries Super-Kamiokande} Collaboration, K.~Abe {\em et~al.}, ``{\em {Search
  for Nucleon Decay via $n \to \bar{\nu} \pi^{0}$ and $p \to \bar{\nu} \pi^{+}$
  in Super-Kamiokande}},''
  \href{http://dx.doi.org/10.1103/PhysRevLett.113.121802}{Phys. Rev. Lett.
  {\bfseries 113} (2014) 121802},
\href{http://arxiv.org/abs/1305.4391}{[{\ttfamily 1305.4391}]}.

\bibitem{Takhistov:2014pfw}
{\bfseries Super-Kamiokande} Collaboration, V.~Takhistov {\em et~al.}, ``{\em
  {Search for Trilepton Nucleon Decay via $p \rightarrow e^+ \nu \nu$ and $p
  \rightarrow \mu^+ \nu \nu$ in the Super-Kamiokande Experiment}},''
  \href{http://dx.doi.org/10.1103/PhysRevLett.113.101801}{Phys. Rev. Lett.
  {\bfseries 113} (2014) 101801},
  \href{http://arxiv.org/abs/1409.1947}{[{\ttfamily 1409.1947}]}.

\bibitem{Abe:2014mwa}
{\bfseries Super-Kamiokande} Collaboration, K.~Abe {\em et~al.}, ``{\em {Search
  for proton decay via $p\to\nu K^+$ using 260 kiloton$\times$year data of
  Super-Kamiokande}},''
  \href{http://dx.doi.org/10.1103/PhysRevD.90.072005}{Phys. Rev. {\bfseries
  D90} (2014) 072005},
\href{http://arxiv.org/abs/1408.1195}{[{\ttfamily 1408.1195}]}.

\bibitem{Fornal:2020bzz}
B.~Fornal, B.~Grinstein, and Y.~Zhao, ``{\em {Dark Matter Capture by Atomic
  Nuclei}},''
\href{http://arxiv.org/abs/2005.04240}{[{\ttfamily 2005.04240}]}.

\bibitem{Zurek:2013wia}
K.~M. Zurek, ``{\em {Asymmetric Dark Matter: Theories, Signatures, and
  Constraints}},'' \href{http://dx.doi.org/10.1016/j.physrep.2013.12.001}{Phys.
  Rept. {\bfseries 537} (2014) 91--121},
\href{http://arxiv.org/abs/1308.0338}{[{\ttfamily 1308.0338}]}.

\bibitem{Aitken:2017wie}
K.~Aitken, D.~McKeen, T.~Neder, and A.~E. Nelson, ``{\em {Baryogenesis from
  Oscillations of Charmed or Beautiful Baryons}},''
  \href{http://dx.doi.org/10.1103/PhysRevD.96.075009}{Phys. Rev. D {\bfseries
  96} (2017) 075009}, \href{http://arxiv.org/abs/1708.01259}{[{\ttfamily
  1708.01259}]}.

\bibitem{Elor:2018twp}
G.~Elor, M.~Escudero, and A.~Nelson, ``{\em {Baryogenesis and Dark Matter from
  $B$ Mesons}},'' \href{http://dx.doi.org/10.1103/PhysRevD.99.035031}{Phys.
  Rev. {\bfseries D99} (2019) 035031},
\href{http://arxiv.org/abs/1810.00880}{[{\ttfamily 1810.00880}]}.

\bibitem{Marraffino:1980dj}
J.~Marraffino, S.~Reucroft, C.~Roos, J.~Waters, M.~Webster, A.~Manz,
  R.~Settles, and G.~Wolf, ``{\em {New Measurement Of $\Sigma$ Decay Properties
  And A Test Of The $|\Delta I| = 1/2$ Rule}},''
  \href{http://dx.doi.org/10.1103/PhysRevD.21.2501}{Phys. Rev. D {\bfseries 21}
  (1980) 2501--2509}.

\bibitem{Zyla:2020zbs}
{\bfseries Particle Data Group} Collaboration, P.~A. Zyla {\em et~al.}, ``{\em
  {Review of Particle Physics}},''
\href{http://dx.doi.org/10.1093/ptep/ptaa104}{PTEP {\bfseries 2020} (2020)
  083C01}.

\bibitem{Eisele:1969fz}
F.~Eisele, R.~Engelmann, H.~Filthuth, W.~Foehlisch, V.~Hepp, E.~Leitner,
  W.~Presser, H.~Schneider, M.~Stevenson, and G.~Zech, ``{\em {Search for
  $\Delta Q=-\Delta S$ Transitions in $\Sigma$-Hyperon Decays}},''
  \href{http://dx.doi.org/10.1007/BF01394070}{Z. Phys. {\bfseries 221} (1969)
  401--410}.

\bibitem{Aaij:2017ddf}
{\bfseries LHCb} Collaboration, R.~Aaij {\em et~al.}, ``{\em {Evidence for the
  rare decay $\Sigma^+ \to p \mu^+ \mu^-$}},''
  \href{http://dx.doi.org/10.1103/PhysRevLett.120.221803}{Phys. Rev. Lett.
  {\bfseries 120} (2018) 221803},
  \href{http://arxiv.org/abs/1712.08606}{[{\ttfamily 1712.08606}]}.

\bibitem{Biagi:1982eu}
S.~Biagi {\em et~al.}, ``{\em {A New Upper Limit For The Branching Ratio
  $(\Xi^- \to n \pi^-) / (\Xi^- \to \Lambda \pi^-)$}},''
  \href{http://dx.doi.org/10.1016/0370-2693(82)90978-9}{Phys. Lett. B
  {\bfseries 112} (1982) 277--280}.

\bibitem{Rajaram:2005bs}
{\bfseries HyperCP} Collaboration, D.~Rajaram {\em et~al.}, ``{\em {Search for
  the lepton-number-violating decay $\Xi^- \to p \mu^- \mu^-$}},''
  \href{http://dx.doi.org/10.1103/PhysRevLett.94.181801}{Phys. Rev. Lett.
  {\bfseries 94} (2005) 181801},
  \href{http://arxiv.org/abs/hep-ex/0505025}{[{\ttfamily hep-ex/0505025}]}.

\bibitem{Shi:2019vus}
X.-D. Shi, X.-W. Kang, I.~Bigi, W.-P. Wang, and H.-P. Peng, ``{\em {Prospects
  for CP and P violation in $\Lambda_{c}^+$ decays at Super Tau Charm
  Facility}},'' \href{http://dx.doi.org/10.1103/PhysRevD.100.113002}{Phys. Rev.
  D {\bfseries 100} (2019) 113002},
  \href{http://arxiv.org/abs/1904.12415}{[{\ttfamily 1904.12415}]}.

\bibitem{Chatrchyan:2013bba}
{\bfseries CMS} Collaboration, S.~Chatrchyan {\em et~al.}, ``{\em {Search for
  Baryon Number Violation in Top-Quark Decays}},''
  \href{http://dx.doi.org/10.1016/j.physletb.2014.02.033}{Phys. Lett.
  {\bfseries B731} (2014) 173--196},
\href{http://arxiv.org/abs/1310.1618}{[{\ttfamily 1310.1618}]}.

\bibitem{Sirunyan:2018xlo}
{\bfseries CMS} Collaboration, A.~M. Sirunyan {\em et~al.}, ``{\em {Search for
  narrow and broad dijet resonances in proton-proton collisions at $
  \sqrt{s}=13 $ TeV and constraints on dark matter mediators and other new
  particles}},'' \href{http://dx.doi.org/10.1007/JHEP08(2018)130}{JHEP
  {\bfseries 08} (2018) 130},
  \href{http://arxiv.org/abs/1806.00843}{[{\ttfamily 1806.00843}]}.

\bibitem{Heeck:2016xwg}
J.~Heeck, ``{\em {Interpretation of Lepton Flavor Violation}},''
  \href{http://dx.doi.org/10.1103/PhysRevD.95.015022}{Phys. Rev. {\bfseries
  D95} (2017) 015022},
\href{http://arxiv.org/abs/1610.07623}{[{\ttfamily 1610.07623}]}.

\bibitem{Hambye:2017qix}
T.~Hambye and J.~Heeck, ``{\em {Proton decay into charged leptons}},''
  \href{http://dx.doi.org/10.1103/PhysRevLett.120.171801}{Phys. Rev. Lett.
  {\bfseries 120} (2018) 171801},
\href{http://arxiv.org/abs/1712.04871}{[{\ttfamily 1712.04871}]}.

\bibitem{Dorsner:2016wpm}
I.~Dor{\v s}ner, S.~Fajfer, A.~Greljo, J.~F. Kamenik, and N.~Ko{\v s}nik,
  ``{\em {Physics of leptoquarks in precision experiments and at particle
  colliders}},'' \href{http://dx.doi.org/10.1016/j.physrep.2016.06.001}{Phys.
  Rept. {\bfseries 641} (2016) 1--68},
\href{http://arxiv.org/abs/1603.04993}{[{\ttfamily 1603.04993}]}.

\bibitem{Araki:2005jt}
{\bfseries KamLAND} Collaboration, T.~Araki {\em et~al.}, ``{\em {Search for
  the invisible decay of neutrons with KamLAND}},''
  \href{http://dx.doi.org/10.1103/PhysRevLett.96.101802}{Phys. Rev. Lett.
  {\bfseries 96} (2006) 101802},
  \href{http://arxiv.org/abs/hep-ex/0512059}{[{\ttfamily hep-ex/0512059}]}.

\bibitem{Barducci:2018rlx}
D.~Barducci, M.~Fabbrichesi, and E.~Gabrielli, ``{\em {Neutral Hadrons
  Disappearing into the Darkness}},''
  \href{http://dx.doi.org/10.1103/PhysRevD.98.035049}{Phys. Rev. D {\bfseries
  98} (2018) 035049}, \href{http://arxiv.org/abs/1806.05678}{[{\ttfamily
  1806.05678}]}.

\bibitem{Takhistov:2015fao}
{\bfseries Super-Kamiokande} Collaboration, V.~Takhistov {\em et~al.}, ``{\em
  {Search for Nucleon and Dinucleon Decays with an Invisible Particle and a
  Charged Lepton in the Final State at the Super-Kamiokande Experiment}},''
  \href{http://dx.doi.org/10.1103/PhysRevLett.115.121803}{Phys. Rev. Lett.
  {\bfseries 115} (2015) 121803},
  \href{http://arxiv.org/abs/1508.05530}{[{\ttfamily 1508.05530}]}.

\end{thebibliography}\endgroup

\end{document}